# Deep Learning for micro-Electrocorticographic (µECoG) Data*

X. Wang, *Student Member*, *IEEE*, C. A. Gkogkidis, R. T. Schirrmeister, F. A. Heilmeyer, M. Gierthmuehlen, F. Kohler *Member*, *IEEE*, M. Schuettler *Member*, *IEEE*, T. Stieglitz, *Senior Member*, *IEEE*, T. Ball

*Abstract* — Machine learning can extract information from neural recordings, e.g., surface EEG, ECoG and µECoG, and therefore plays an important role in many research and clinical applications. Deep learning with artificial neural networks has recently seen increasing attention as a new approach in brain signal decoding. Here, we apply a deep learning approach using convolutional neural networks to µECoG data obtained with a wireless, chronically implanted system in an ovine animal model. Regularized linear discriminant analysis (rLDA), a filter bank component spatial pattern (FBCSP) algorithm and convolutional neural networks (ConvNets) were applied to auditory evoked responses captured by µECoG. We show that compared with rLDA and FBCSP, significantly higher decoding accuracy can be obtained by ConvNets trained in an end-to-end manner, i.e., without any predefined signal features. Deep learning thus proves a promising technique for µECoG-based brain-machine interfacing applications.

## I. INTRODUCTION

Recording techniques with geometries (contact diameter and inter-contact distance) smaller than those of conventional clinical electrode arrays (contact diameter with several mm and an inter-contact distance on the order of 1 cm; [1]) may collectively be defined as microelectrocorticography (µECoG). µECoG has high temporal and spatial resolution, and is less invasive and may decrease the surgery risk as well as complications after surgery due to its compact geometry [2-3]. A previous study using µECoG in a minipig model [4] showed that spatial information can be obtained in higher detail using finer array geometry and smaller contact diameters, and that certain high-frequency signal components may be better captured.

*Research supported by German Federal Ministry of Education and Research (BMBF grant 13GW0053D MotorBic to BFNT Freiburg/ Tübingen, BMBF grant 0316064C BrainCon) and Deutsche Forschungsgemeinschaft (DFG grant EXC1086 BrainLinks-BrainTools (BLBT)) to the University of Freiburg, Germany.

X.W. and C.A.G. are with Translational Neurotechnology Lab, Department of Neurosurgery, Faculty of Medicine, Medical Center - University of Freiburg, University of Freiburg, Germany and with Laboratory for Biomedical Microtechnology, Department of Microsystems Engineering (IMTEK), University of Freiburg, Germany

M.G. is with Department of Neurosurgery, University Hospital Knappschaftskrankenhaus Bochum, Ruhr-University Bochum, Germany. This author's current address is different from the address where the work was carried out.

F.K. and M.S. are associated to CorTec GmbH, Freiburg, Germany.

T.S. is with Laboratory for Biomedical Microtechnology, Department of Microsystems Engineering (IMTEK), University of Freiburg, Germany

R.T.S., F.A.H. and T.B. are with Translational Neurotechnology Lab, Department of Neurosurgery, Faculty of Medicine, Medical Center - University of Freiburg, University of Freiburg, Germany

(Corresponding author phone: +49-761-270-87570; e-mail: xi.wang@uniklinik-freiburg.de).
.

Machine learning can be used to extract information from neural signals for neurotechnological applications. There are various machine learning methods being developed for this end. For example, regularized linear discriminant analysis (rLDA) computes an explicit, simple relationship from the raw data [5]; filter bank component spatial pattern (FBCSP) algorithms are frequently applied to data obtained with intracranial recordings, which compute hand-crafted features using prior knowledge about the domain [6]; deep learning with convolutional neural networks (ConvNets) is a recently very popular and successful machine learning method which learns features for which the range of learnable features is only implicitly constrained (i.e., neural networks only by network architecture; see [7-8] for further explanations).

rLDA has been shown to be able to successfully distinguish different grasp movements in human ECoG data, for example in [9-10]. The FBCSP algorithm is an effective method for classifying 2-class motor imagery in surface electroencephalographic (EEG) data [11]. ConvNets applied to EEG data of healthy human subjects yielding performance at least as good as other widely used decoding algorithms, but without prior decisions regarding the signal features to be used [12]. The decoding algorithms mentioned above have, to our knowledge, never been compared on the basis of µECoG recordings. The aim of the present study was to evaluate the feasibility of using deep learning with ConvNets for µECoG recordings obtained from sheep auditory cortex in a long term implantation setting.

## II. MATERIALS & METHODS

### A. Experimental parameters

One sheep (ovis orientalis aries, Fig. 1a) was chronically implanted with a µECoG-based neural interfacing device (Fig. 1a). Electrode arrays (Fig. 1b) were placed subdurally on the motor and auditory cortex (Fig. 1c, d) and a hermetic package carrying the electronics was located in the back of the sheep (Fig. 1a). The animal was implanted for 162 days. Auditory stimuli were presented on 15 experiment days and consisted of three seconds of bandpass-filtered white noise (8 kHz center frequency, 125 kHz bandwidth) followed by a two-second pause. Stimuli were presented with a loudspeaker centered 70 cm in front of the sheep's head. Loudness was calibrated to an average of 75.8 dB SPL (±1.7 dB SPL near the sheep's ear) using a sound level meter (Model AL1 Acoustilyzer, NTi Audio AG, Schaan, Liechtenstein). This five-second trial was repeated for 261 times on average (mean across all experiment days) on each experiment day.

For details on the chronic implantation surgery and anesthesia medication see [13]. All experiments in the present study were approved by the Animal Committee of the University of Freiburg and the Regierungspräsidium Freiburg,

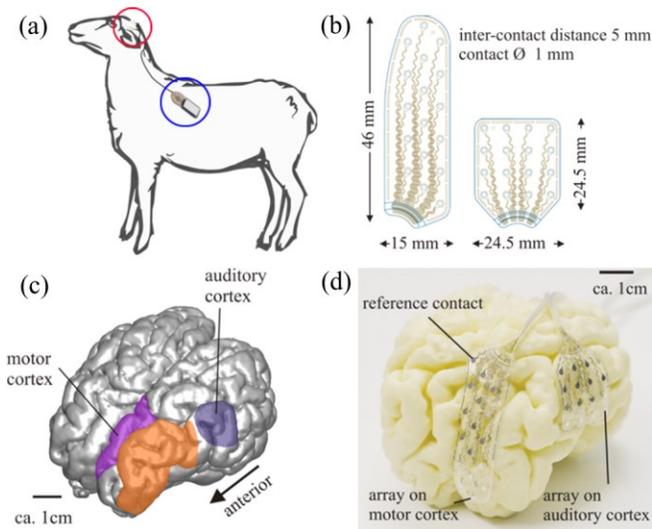

Fig. 1. Chronic experimental setup and µECoG implant. (a) Schematic illustration of µECoG implant position with two µECoG electrode arrays (inside red circle) and a hermetic package (inside blue circle; amplifiers, infrared transmission system, inductive power supply, etc.) and position of animal during chronic experiments. (b) Blueprint of µECoG electrode arrays. The left-hand array was implanted on motor cortex (purple area in (c)) and the right-hand array on auditory cortex (blue region in (c)). (c) MRI reconstruction of a sheep brain. This figure is modified from [14]. d) MRI-based 3D-printed sheep brain with the approximate location of the µECoG electrode arrays on the motor and auditory cortex as shown in (c). (b) and (d) are originated from CorTec GmbH, Freiburg, Germany.

Baden-Wuerttemberg and EU Directive 2010/63/EU was followed for all experiments.

### B. µECoG implant

The µECoG-based implant used in this study was the *Brain-Interchange system* (designed and manufactured by CorTec GmbH, Freiburg, Germany; [15]). This is an active implantable device that receives power from and communicates with an external unit using a wireless connection. Electrode arrays and a hermetic electronic package are the main units for the implant system and are connected by a 50 cm cable (Fig. 1a)

The implant recorded neural signals from two separate electrode arrays (Fig. 1b). Each array consisted of 16 platinum-iridium electrode contacts (Ø 1 mm, inter-contact distance 5 mm) which were manufactured in a customizable laser-based fabrication process [16]. Neural signals were recorded at a sampling rate of 900 Hz.

The implant was inductively powered by the extracorporeal transceiver unit, at a carrier frequency of 250 kHz. Communication to the extracorporeal transceiver unit was effected via infrared link; data storage was achieved via direct connection to the computer using a USB cable. An external trigger signal was fed in to the external unit to allow for precise alignment of brain responses with the stimuli. Amplifiers and infrared telemetry were hermetically sealed using a double-sealing methodology in order to improve long-term reliability of signal quality [17].

### C. Data preprocessing

The main neural responses analyzed in this study were the auditory evoked potentials (AEPs) and associated changes in spectral power elicited by auditory stimulation. Thus, only the data recorded by the 16 contacts from the electrode array located on the auditory cortex were included in the analyses described here (i.e., the right-hand side array in Fig. 1b, d). Prior to more specific processing, all recorded data were re-referenced to a common average reference of all 16 contacts. After re-referencing, noisy contacts (defined as 20% of the data exceeding 800 µV) were excluded for further analysis. Raw data from the remaining contacts were re-referenced again and high-pass filtered (cut-off frequency 0.5 Hz, second-order Butterworth high-pass filter) for a general overview and for spectral power computation. For decoding, high-pass filtered data were additionally low-pass filtered (cut-off frequency 120 Hz, second-order Butterworth low-pass filter).

The single trials of the µECoG recordings comprised three stages (Fig. 2): 1 s before stimulus onset, 3 s during the auditory stimulation period and 1 s after stimulus offset. Each five-second stimulus-related response was averaged for each contact and each experimental day. During the early response period (first 20 - 200 ms after the stimulus onset), maps of AEP amplitudes were computed for all contacts to visualize topographic distribution of the recorded signals.

### D. Spectral power analysis

High-pass filtered data (see above) were pre-whitened before performing time-resolved Fast Fourier Transform (FFT). These 5-s-stimulus related responses were analyzed with a sliding window of 250 ms and a window step of 80 ms resulting in 131 frequency bins (0 - 450 Hz in steps of 4 Hz) and 60 time bins. Relative spectral power changes (relSPs) were computed by dividing absolute spectral power by baseline spectral power (average across 10 time bins before stimulus onset).

### E. Class definition

Our main analysis was intended to detect and decode brain responses during different phases of auditory stimulation, including the difference between early and late responses post-stimulus onset. Decoding information from auditory cortex could in future be useful for closed-loop control of cochlear implants (see Discussion). Based on the stimulus onset, we separated the 5 s stimulus-related responses into five epochs of 1 s, and assigned them to one of two types of decoding analysis: (1) 2-class datasets, for signals with or without auditory stimulation (as shown in Fig. 2a), and (2) 3-class datasets for early (first second after stimulus onset) and later (second and third seconds after stimulus onset) auditory related responses classified as 'Response 1', 'Response 2' or 'Response 3' (see Fig. 2b). By assigning the recorded data from each experiment day into different classes and labeling them separately as described above, we thus obtained different trials of different classes ('class-trials' defined as 1s-epoch responses with assigned class labels), which constituted the input data to the following decoding process.

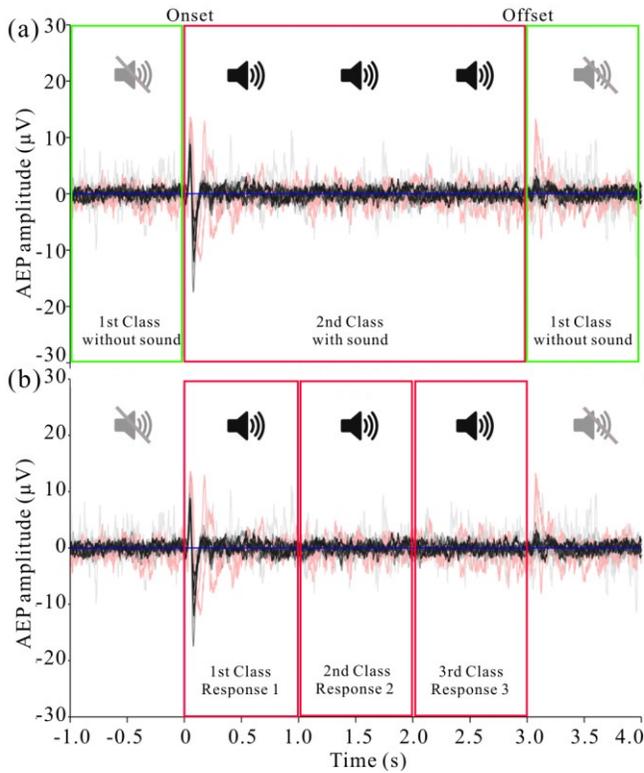

Fig. 2. Class assignment for µECoG recordings of auditory evoked potentials (AEPs). This is an example of the averaged 5-s stimulus-related responses from one electrode contact from individual experiment days. Black and grey traces depicts measurements in awake condition (n=13), red traces depict measurements under general anesthesia (n=2). Green and red boxes in (a) indicate 2-class datasets defined for the first decoding task, namely to distinguish neural responses with or without auditory stimulation. The three red boxes in (b) delineate the 3-class datasets as defined for the second decoding task, intended to distinguish early and late responses during auditory stimulation.

*F. Decoding analysis*

In order to make decoding results from the three methods under consideration comparable, the overall decoding architecture was kept consistent, as shown in Fig. 3a. Preprocessed input data was divided into three parts, as follows: training data set (the first 64% of the input data, or 80% of the first 80%) was used to build a detection model. To optimize the parameters of the detection model, detection performance was evaluated on the validation data set for different values of the parameters (a further 16% of the input data: the remaining 20% of the first 80%). The parameters yielding the best performance were then used to construct the classifying model. The last 20% of the total data were then used to evaluate the performance of this model. Comparison of decoded labels and assigned labels from the validation datasets was used to evaluate overall detection performance; better detection performance was defined as a higher similarity between the two sets of labels.

Regularized linear discriminate analysis (rLDA) and the filter bank component spatial pattern algorithm (FBCSP) are feature extraction methods which find a linear/non-linear combination of features which separate two or more classes of events. Deep learning with convolutional neural networks (ConvNets) is a non-linear method which processes the input signal through several layers, where intuitively high-level, more abstract concepts can be learned from lower-level and more concrete concepts (see [7-8] for details). The main difference between the three methods was how to build the detection/classification model in the training phase, as shown in Fig. 3b. In the following we explain the steps of the construction of the detection model for each method separately.

*Decoding with rLDA:*

- Class-trials which included data samples larger than 800 µV were defined as 'bad trials'. When located in the training or validation part, these were excluded from further analysis. However, 'bad trials' in the final testing dataset were kept. This was in order to construct a model which can handle the robust data more efficiently, as it is more similar to a 'real-life' online decoding situation.
- The rLDA detection model assumes a Gaussian distribution of the cleaned class-trials in the training datasets for each class, and returns the probability of belonging to one of the classes for any arbitrary point in the feature space. The fitted Gaussian distributions share a common covariance matrix for all classes (see [5, 10] for details).

*Decoding with FBCSP:*

- Bad trials were removed according to the same criteria as with rLDA
- Spatial filtering was performed using the common spatial pattern (CSP) algorithm, which linearly transforms the input data (see [6, 11] for details)
- Log-variance of the spatially filtered class-trails was calculated to create the feature vectors for each frequency band

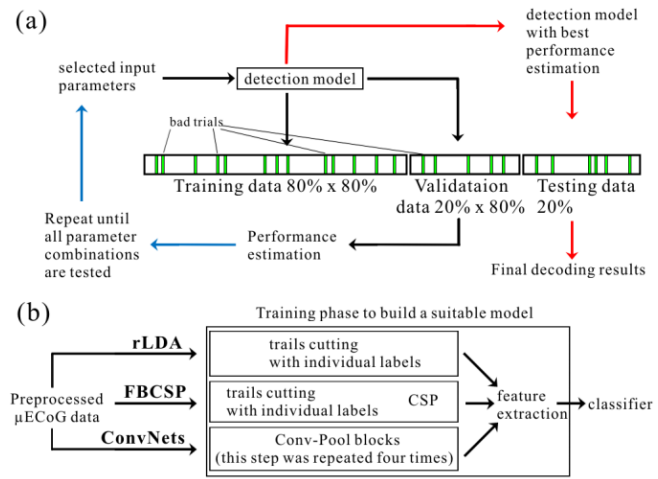

Fig. 3. Architecture of decoding process for the three methods addressed in the present study. (a) Input datasets were divided into three parts. The first 80% of the input data was used to compute a classifying model, of which 80% was used for training and 20% for validation. The remaining 20% of the input data was used for testing. Bad class-trials (green blocks) were excluded from the training and validation datasets during the decoding process. (b) Steps involved in the construction of classifying models in the training phase, for each of the three decoding methods addressed in present study.

*Decoding with ConvNets:*

- Bad trials were removed according to the same criteria as with rLDA
- Our ConvNets implementation used four convolution-max-pooling blocks (Conv-Pool blocks), with a special first block designed for large input datasets; this block had two layers in order to separate the overall convolution into a linear combination of temporal convolution and spatial filtering. This was followed by three standard Conv-Pool blocks (temporal convolution intended to extract the features and max-pooling to decrease the size of input data without losing feature information for further computation) and a dense softmax classification (see [12] for details).
- To improve performance, we used exponential linear units (ELUs, $f(x) = x$ for $x > 0$ and $f(x) = e^x - 1$ for $x <= 0$) as activation functions for each layer output; batch normalization to facilitate the optimization, and applied random dropout to each layer to prevent co-adaption and over-fitting (see [12] for details)
- Optimization used stochastic gradient descent with the Adam optimizer with a learning rate of 0.1%, a batch size of 32 samples up to 500 epochs (i.e., class-trials) with early stopping as described in [12]

Next, Confusion matrices for all three decoding methods were computed, which were $(n+1)$ x $(n+1)$ ($n$ = number of class) squares color-coded with specific numbers to allow visualization of the performance of each decoding method ([18]; Tab. 1). Information in each small square at row $r$ and column $c$ was the number of trials of target $r$ predicted as class $c$, i.e., $D_{rc}$ in Table 1, also written as a percentage of all trials in the test dataset,

$$\frac{D_{rc}}{\sum_{r=1}^{n}\sum_{c}^{n} D_{rc}}, r = 1:n, c = 1:n$$

this percentage value in the diagonal direction stands for the decoding accuracy (DA) for each class,

$$DA_i = \frac{D_{ii}}{\sum_{r=1}^{n}\sum_{c=1}^{n} D_{rc}}, i = 1:n$$

And the final DA over all class is equal to,

$$DA = \frac{\sum_{r,c=1}^{n} D_{rc}, \ r = c}{\sum_{r=1}^{n}\sum_{c=1}^{n} D_{rc}}$$

Information in the $n+1$ column gives the precision, defined as the proportion of target trials correctly predicted for class $r$ as class $c$,

$$P_r = \frac{D_{rr}}{\sum_{c=1}^{n} D_{rc}}, r = 1:n$$

Information in the $n+1$ row gives the sensitivity, defined as the proportion of target trials correctly predicted for class $c$ as class $r$,

$$S_c = \frac{D_{cc}}{\sum_{r=1}^{n} D_{rc}}, c = 1:n$$

Statistical comparison (by wilcoxon-ranksum test) was also performed on all information obtained for our three decoding methods and presented in the confusion matrices.

## III. RESULTS

### A. AEPs and topographic maps

Fig. 4a shows the auditory responses from the averaged 5-s stimulus-related responses from one electrode contact on each experiment day, and Fig. 4b shows the topographic maps over all contacts for the early part of the response, on certain experiment days. Together, the time-domain signal components, signal strength, and topographic distribution all suggest that the chronically implanted µECoG array reliably recorded the neural response elicited by acoustic stimulation.

### B. Spectral power

Time-frequency analysis showed that acoustic stimuli elicited robust auditory evoked responses, peri-stimulus spectral power increase in the 5 - 40 Hz band, and post-stimulus spectral power increase in the 50 - 150 Hz band across the whole implantation period (Fig. 5).

Table 1. Information structure of confusion matrices

|         |         | Predicted |         |     |         | precision |
|---------|---------|-----------|---------|-----|---------|-----------|
|         |         | Class 1   | Class 2 | ... | Class n |           |
| Actual  | Class 1 | $D_{11}$  | $D_{12}$ | ... | $D_{1n}$ | $P_r = \frac{D_{rr}}{\sum_{c=1}^{n} D_{rc}}$, $r = 1:n$ |
|         | Class 2 | $D_{21}$  | $D_{22}$ | ... | $D_{2n}$ |           |
|         | ...     | ...       | ...     | ... | ...     |           |
|         | Class n | $D_{n1}$  | $D_{n2}$ | ... | $D_{nn}$ |           |
| sensitivity |     | $S_c = \frac{D_{cc}}{\sum_{r=1}^{n} D_{rc}}$, $c = 1:n$ | | | | $DA = \frac{\sum_{i=1}^{n} D_{ii}}{\sum_{r=1}^{n}\sum_{c=1}^{n} D_{rc}}$ |

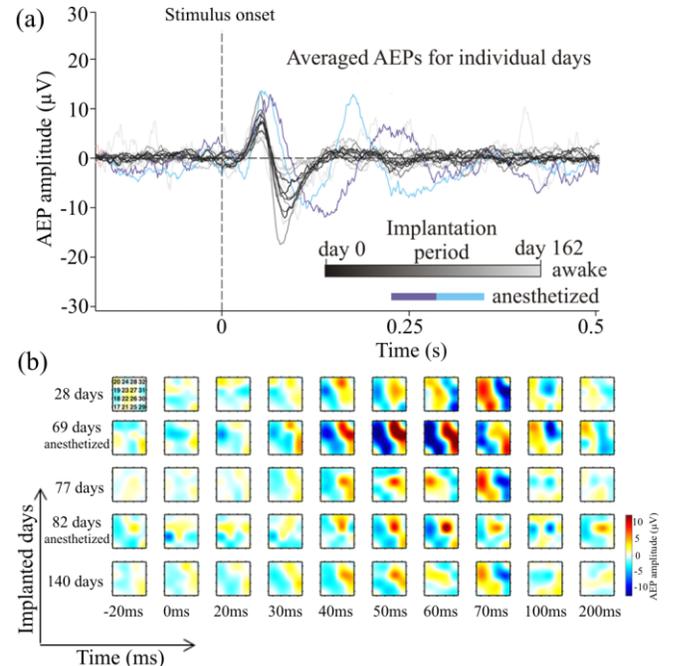

Fig. 4. Averaged auditory evoked potentials (AEPs) and topographic maps from certain experiment days. (a) Averaged AEP response from one contact over auditory cortex from on 15 different experiment days. Dark to light gray color of the AEP traces depict earlier to later measurements during the implantation period under awake condition, and dark to light blue traces depict averaged AEPs from earlier to later experiment days under anesthesia. This figure is modified from [15]. (b) Topographic AEP maps obtained from all contacts for five selected experiment days. Each panel represents the µECoG array over auditory cortex. Warm to cold colors indicate positive to negative AEP values.

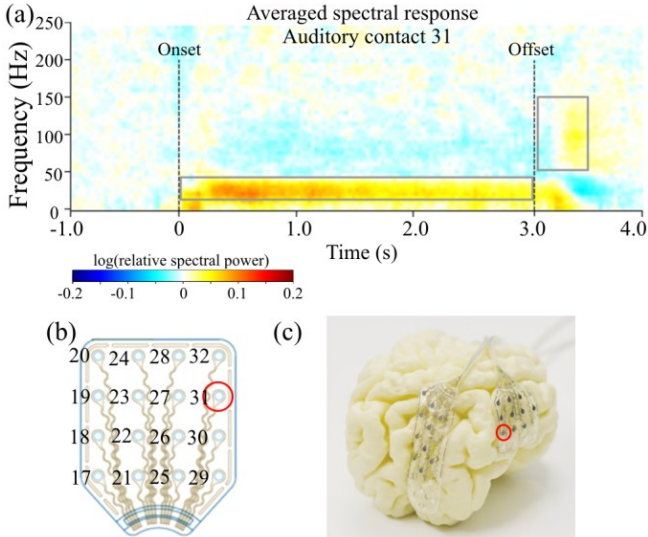

Fig. 5. Averaged auditory relative spectral response from one auditory contact (b, c) across all 13 experiment days for measurements ender awake conditions. Warm colors indicate increase in spectral power and cold colors indicate decrease in spectral power compared to baseline. (b) and (c) are originated from CorTec GmbH, Freiburg, Germany.

*C. Decoding accuracies*

Decoding accuracy (DA) as reported here describes the proportion of correctly labeled class-trials in the test dataset. Statistical comparison (wilcoxon-ranksum test between decoding methods) shoed, for the 2-class input dataset, that DAs with ConvNets are higher that rLDA (Fig. 6a, left, $p < 0.01$). For the 3-class input dataset, DAs with ConvNets are significantly higher than FBCSP (Fig. 6b, right, $p < 0.001$). There is no significant difference in DA between ConvNets and FBCSP for 2-class input datasets (Fig. 6a, right). There is no significant difference in DA for 3-class input datasets between ConvNets and rLDA (Fig. 6b, left). However, DA for 3-class input datasets with ConvNets is similar to that obtained with rLDA (linear correlation coefficient: 0.658).

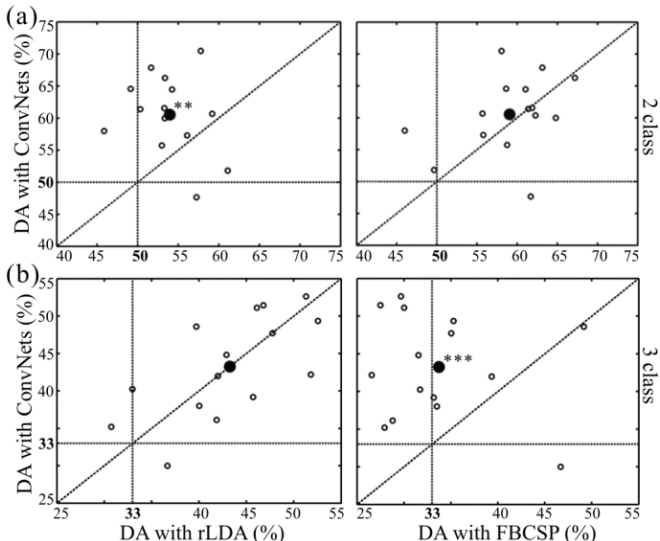

Fig. 6. Comparison of decoding accuracies (DAs) with rLDA and FBCSP versus ConvNets, for both the 2-class and 3-class decoding process. Each small open circle represents the DA from one experiment day, while large closed circles represent the average DA over all experiment days. Circles above the dotted diagonal line indicate decoding process where ConvNets performed better than rLDA (a, left for 2-calss; b, left for 3-class) or FBCSP (a, right for 2-calss; b, right for 3-class). The opposite holds of circles below the dashed line. Stars indicate statistically significant differences between ConvNets and rLDA or between ConvNets and FBCSP (wilcoxon-ranksum test; $p < 0.05$: *, $p < 0.01$: **, $p < 0.001$: ***). Horizontal and vertical dotted lines indicate the chance level (50% for 2-calss and 33% for 3-class) for decoding detection.

*D. Confusion matrices*

Confusion matrices for 2-class input dataset are very similar for ConvNets, rLDA and FBCSP (Fig. 7, upper row). As explained before how the confusion matrix was computed (Tab. 1), the precision values are significantly different for the comparison between all three decoding methods (wilcoxon-ranksum test). Confusion matrices for the 3-class input dataset are only similar for ConvNets and rLDA (Fig. 7, lower row). The values suggest that if we merge 'response 2' and 'response 3' into a single class, then decoding with ConvNets and rLDA may successfully distinguish the 'response 1' from this new class. This would be consistent with the input data structure of the 3-class datasets, since the main early neural responses were elicited during the 'response 1' period (Fig. 2b). However, confusion matrices with FBCSP for the same input data showed greater variability and hence a less consistent pattern.

## IV. DISCUSSION & CONCLUSION

We have reported what is to our knowledge for the first application of deep learning to µECoG data. We analyzed auditory-evoked cortical brain responses obtained with a chronically implanted, wireless neural interface system and presented detailed results on how, compared to rLDA or FBCSP, ConvNets produced better decoding performance.

Decoding auditory information from cortical areas could have applications in cochlear implants (CIs) controlled by brain-computer interfacing: For example, CIs at present cannot reproduce the so called "cocktail-party capability" of healthy listeners, i.e., filtering individual speakers in noisy environments through selective auditory attention. Decoding auditory targets from brain activity could provide control signals for online adaptation of CI filter settings, in order to restore selective auditory attention. However, this would require decoding of fine-grained attention-related signals. We anticipate that this task will require powerful machine-learning algorithms. Our findings here suggest that deep learning is a promising candidate for such applications.

For the further study, we perform cross-validation (CV) during model construction, in order to increase the input data sample size. CV might influence the decision on how much the decoded DAs will differ from chance level, which is theoretically valid for infinite sample size. One limitation of the present study is that many other machine learning algorithms are commonly applied in neural signal classification, e.g., naïve Bayers (NB), support vector machines (SVMs), etc. More comprehensive comparisons to other machine learning algorithms incorporating our µECoG recordings are needed to underpin the strength of deep learning approaches and are currently under investigation. In addition, other deep learning (-related) techniques, such as recurrent or residual networks, data augmentation, automatic network architecture search etc., should also be included and assessed in comparative studies. Further, suitable visualization techniques can provide insights into what the networks learn from the data [12, 19]. Our present results provide encouragement that the full potential of deep learning will soon be brought to bear on μECoG-based neurotechnology.

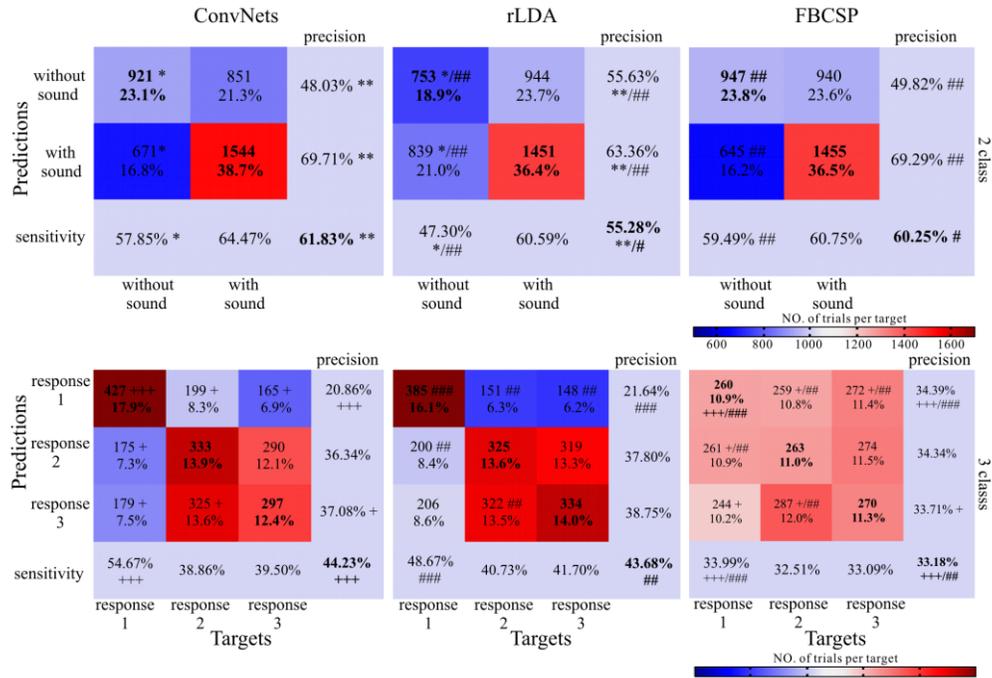

Fig. 7. Confusion matrices for decoding processes using rLDA, FBCSP and ConvNets. Results are shown for both 2-class and 3-class decoding processes, summarized over all 15 experiment days. Information in each panel (row *r*, column *c*) from upper-left 2-by-2 or 3-by-3 squares: number of trials of target *r* predicted as class *c* (also written as a percentage of all trials in test dataset). Information in the bold-font panels along the diagonal corresponds to correctly predicted trials of the different classes. Percentages and colors indicate proportion of trials in this panel within all trials of the corresponding column (i.e., from all trials of the corresponding target class). The lower-right value represents overall accuracy. The bottom row represents sensitivity, defined as the number of trials correctly predicted for class *c*/number of trials for class *c*. The rightmost column represents precision, defined as the number of trials correctly predicted for class *r*/number of trials predicted as class *r*. Asterisks indicate statistically significantly different decoding results between ConvNet and rLDA, Hashmarks (#) markers indicate comparison between rLDA and FBCSP, and plus signs (+) comparison between ConvNets and FBCSP ($p < 0.05$: */#/+, $p < 0.01$: **/##/++, $p < 0.001$: ***/###/+++, wilcoxon-ranksum test).

ACKNOWLEDGMENT

The author would like to thank Dr. T. Freimann for assisting with the surgical component of the animal experiments; K. Foerster and Prof. J. Haberstroh for maintaining the anesthesia during the surgery.

FK and MS are affiliated with company CorTec GmbH which developed the wireless neural implant system in this study. Other authors do not have any affiliation with CorTec, and they do not have any other conflict of interest to declare.

REFERENCES

[1] J. Jr. Engel, "Surgery for seizures," *New Engl. J. Med.*, vol. 334, pp 647-653, 1996.
[2] H. M. Hamer et al, "Complications of invasive video-EEG monitoring with subdural grid electrodes," *Neurology*, vol. 58, no. 1, pp. 97-103, 2002.
[3] C. H. Wong et al, "Risk factors for omplications during intracranial electrode recording in presurgical evaluation of drug resistant partial epilepsy," *Acta. Neurochir.*, vol. 151, pp. 37-50, 2009.
[4] X. Wang, C.A. Gkogkidis, et al, "Mapping the fine structure of cortical activity with different micro-ECoG electrode array geometries," *J. Neural Eng.*, vol. 14, no. 5, pp. 056004, Jun 2017.
[5] J. H., Friedman, "Regularized Discriminant Analysis," *Journal of the American Statistical Association*, vol. 84, pp. 165-175, 1989.
[6] Z. Y., Chin, K. K., Ang, C., Wang, C, Guan and H., Zhang, "Multi-class filter bank common spatial pattern for four-class motor imagery BCI," *Conf. Proc. IEEE Eng. Med. Biol. Soc.*: 571-574, Sep 2009.
[7] Y. LeCun, Y. Bengio and G. Hinton, "Deep learning," *Nature*, vol. 521, no. 7553, pp. 436-444, May 2015.
[8] J. Schmidhuber, "Deep learning in neural networks: an overview," *Neural Netw.*, 61: 85-117, Jan 2015.
[9] T. Pistohl, A. Schulze-Bonhage, A. Aertsen, C. Mehring and T. Ball, "Decoding natural grasp types from human ECoG," *Neuroimage*, vol. 59, no. 1, pp. 248-260, Jan 2012.
[10] T. Milekovic, T. Ball, A. Schulze-Bonhage, A. Aertsen and C. Mehring, "Detection of error related neuronal responses recorded by electrocorticography in humans during continuous movements," *PLoS One*, vol. 8, no. 1, pp. e55235, 2013.
[11] K. K., Ang, Y., Zheng, C., Wang, C. G and Haihong Zhang, "Filter Bank Common Spatial Pattern Algorithm on BCI Competition IV Datasets 2a and 2b", *Front Neurosci.*, 6: 39.
[12] R. T. Schirrmeister et al, "Deep learning with convolutional neural networks for EEG decoding and visualization," *Hum. Brain Mapp.*, vol. 38, no. 11, pp. 5391-5420, Nov 2017.
[13] M. Gierthmuehlen, X. Wang, C.A. Gkogkidis, et al, "Mapping of sheep sensory cortex with a novel microelectrocorticography grid," *J. Comp. Neurol.*, vol. 522, no. 16, pp 3590-3608, 2014.
[14] C.A. Gkogkidis, X. Wang, et al, "Closed-loop interaction with the cerebral cortex using a novel micro-ECoG-based implant: The impact of beta vs. gamma stimulation frequencies on cortico-cortical spectral responses," *Brain-Computer Interfaces*, vol. 4, no. 4, pp. 214-24, 2017.
[15] F. Kohler, C. A., Gkogkidis et al, "Closed-loop interaction with the cerebral cortex: A review of wireless implant technology," *Brain-Computer Interfaces*, vol. 4, no. 4, pp. 146-154, 2017.
[16] M. Schuettler, S. Stiess, B. V. King and G. J. Suaning, "Fabrication of implantable microelectrode arrays by laser cutting of silicone rubber and platinum foil," *J. Neural Eng.*, vol. 2, no. 1, pp. S121, 2005.
[17] M. Schuettler, F. Kohler, J. S. Ordonez and T. Stieglitz, "Hermetic electronic packaging of an implantable brain-machine-interface with transcutaneous optical data communication," *Conf. Proc. IEEE Eng. Med. Biol. Soc.*, pp. 3886-3889, 2012.
[18] R. Kohavi and F. Provost, "Glossary of Terms," *Machine Learning*, 30: 2-3, 1998
[19] P. J. Kindermans et al, "PatternNet and PatternLRP - Improving the interpretability of neural networks," *CoRR.*, vol. abs/1705.05598, 2017.